\documentclass[%
 aps,
 prb,%
 amsmath,amssymb,
floatfix,
reprint,%
showpacs,
]{revtex4-1}

\pdfoutput=1

\usepackage{inputenc}
\usepackage{graphicx}
\usepackage{siunitx,booktabs}
\usepackage{url}
\usepackage{dcolumn}
\usepackage{color}
\usepackage{multirow}
\usepackage[american]{babel}
\usepackage{setspace}

\newcolumntype{L}[1]{>{\raggedright\arraybackslash}p{#1}}
\newcolumntype{C}[1]{>{\centering\arraybackslash}p{#1}}
\newcolumntype{R}[1]{>{\raggedleft\arraybackslash}p{#1}}

\newcommand{\GW}[1]{$G_0W_0$#1}
\newcommand{\Ref}[2]{\textit{#1}: Ref. [\onlinecite{#2}]}
\newcommand{\exciting}{{\usefont{T1}{lmtt}{b}{n}exciting}}
\newcommand{\rev}[1]{{\color{black} #1}}

\begin{document}

\def\ITA{Grupo de Materiais Semicondutores e Nanotecnologia, Departamento de F\'{i}sica, Instituto Tecnol\'{o}gico de Aeron\'{a}utica, 12228-900 S\~{a}o Jos\'{e} dos Campos, SP, Brazil}
\def\HU{Physics Department and IRIS Adlershof, Humboldt-Universit\"at zu Berlin, Zum Gro{\ss}en Windkanal 6, D-12489 Berlin and European Theoretical Spectroscopy Facility (ETSF)}

\title{
Probing the LDA-1/2 method as a starting point for \GW{} calculations
}

\author{Ronaldo Rodrigues Pela} \email{ronaldorpela@gmail.com} \affiliation{\HU}\affiliation{\ITA} 
\author{Ute Werner, Dmitrii Nabok, Claudia Draxl}\affiliation{\HU}

\date{\today}

\begin{abstract}
\rev{Employing the \GW{} approximation of Hedin's $GW$ approach one can obtain quasi-particle energies of extended systems and molecules with good accuracy.} However, for many materials, semi-local exchange-correlation functionals are unsatisfactory starting points for \GW{} calculations. Hybrid functionals often improve upon them, but at a \rev{substantially} higher computational cost. As an alternative, we suggest the LDA-1/2 method, which provides reasonable \rev{band gaps}, without being computationally involved. In this work, we systematically compare 3 starting points for $G_0W_0$: LDA, PBE0, and LDA-1/2. A selection of solids is chosen for this benchmark: C, Si, SiC, AlP, LiF, MgO, Ne, Ar, GaN, GaAs, CdS, ZnS, and ZnO. We demonstrate that LDA-1/2 is a good starting point in most cases, reducing the mean absolute error \rev{of band gaps} by 50\% when compared to the other 2 functionals. 
\end{abstract}

\maketitle

\section{Introduction}

\par A major goal of first-principles calculations is to reliably predict the properties of materials, in order to guide the design of new materials, or to better understand the behavior of those already available. Describing single-particle excitations well is among the key elements when it comes to materials for optoelectronic applications. The more accurately these properties are obtained from first-principles calculations, the better these calculations can explain experiments, provide insight, or propose new materials \rev{with improved properties.}\cite{Gallandi2016,Koerzdoerfer2012} The $GW$ approach\cite{Hedin1965} of many-body perturbation theory (MBPT) has become a standard framework to calculate single-particle excitations in solids.\cite{Aryasetiawan1998,Onida2002,Nabok2016} In its original formulation by Hedin, a set of coupled integro-differential equations needs to be solved self-consistently to determine the single-particle \rev{Green function,} and, from its poles, the single-particle excitation energies. On the other hand, in the more pragmatic single-shot approximation, \GW{}, the self-consistency is abandoned and Hedin's equations are solved up to first iteration. In this case, a well-defined set of wavefunctions and eigenvalues is employed as the starting point, in order to obtain a new set of eigenvalues including quasi-particle (QP) corrections. 


\par The simplest starting point from Kohn-Sham (KS) calculations is the local density approximation (LDA). QP energies obtained by \GW{} on top of LDA (\GW{@}LDA) are in good agreement with experiments, especially for $sp^3$ bonded materials.\cite{Aryasetiawan1998,Schilfgaarde2007} Nevertheless, particularly for materials containing $d$ or $f$ electrons, LDA turns out to be an inadequate starting point. For such materials \GW{@}LDA fails (even qualitatively) due to the intrinsic lack of localization of LDA wavefunctions. \cite{Schilfgaarde2007,Schilfgaarde2006a,Walsh} Overall, despite the success of \GW{@}LDA, this approximation underestimates band gaps by a slight amount in some cases, but substantially in others.\cite{Schilfgaarde2006b} In fact, only if the underlying KS results are close to the quasiparticle eigenvalues, the perturbative treatment is well justified, and often this is not the case for semi-local exchange-correlation (XC) functionals.

\par Self-consistent $GW$ is expected to be more accurate and independent of the starting point, as in the original formulation of Hedin.\cite{Hedin1965} Obviously, it is computationally much more involved. An alternative is to remain in the framework of \GW{}, but to change to a better starting point.\cite{Koerzdoerfer2012,Marom2012,Bruneval2013,Atalla2013} An effective choice is often presented by hybrid functionals, which combine a fraction of Hartree-Fock exchange with semi-local functionals. For instance, \GW{} evaluated on top of hybrid functionals improves upon \GW{@}LDA \rev{in terms of accuracy.}\cite{Fuchs2007,Bruneval2013,Marom2012} A major disadvantage, however, is the computational cost, as hybrid functionals may be two orders of magnitude more expensive than LDA.\cite{Lucero2012,Pela2015} An additional drawback is that there is no universal mixing parameter for the fraction of the Hartree-Fock \rev{exchange. Although} some systematic ways of finding it have been proposed,\cite{Chen2014,Marques2011,Walsh,Alkauskas2008,Alkauskas2011,Koerzdoerfer2012,Atalla2013,Atalla2016} they are still material dependent to some extent.\cite{Perdew1996,Alkauskas2008,Komsa2010,Marques2011,Alkauskas2011}

\par In this manuscript, we address the issue of a good and at the same time efficient starting point for \GW{} and evaluate the LDA-1/2 method for this purpose.\cite{Ferreira2008,Ferreira2011,Ferreira2012} It implements Slater's transition state technique\cite{Adachi,Slater1971,Slater1972a,Slater1972b} for solids. The LDA-1/2 method improves (over LDA) in terms of band gaps, band alignments, and defect levels. \cite{Ferreira2012,Pela2011,Pela2012,Filho2013,Matusalem2013,Matusalem2014} To assess whether the LDA-1/2 method is a suitable starting point for \GW{} calculations, we compare it with LDA and PBE0. For a series of selected solids, we study band gaps, the position of $d$ levels, and band structures, confronting \GW{} calculations, based on these three distinct starting points, \rev{with} each other and with experimental data.

\section{Methods}\label{sec-methods}
\par While PBE0 requires a generalized KS (gKS) scheme,\cite{Seidl1996,Perdew1996} LDA-1/2 stays within the local KS framework. It is based \rev{on Janak's theorem,}\cite{Janak1978} and on the assumption of a linear behavior of the KS eigenvalues with \rev{their} occupation.\cite{Leite1971} With these premises, the ionization energy $I$ can be expressed as the \rev{negative highest} occupied KS eigenvalue, $-\varepsilon_{v}(1/2)$, with half-occupation. Accordingly, the electron affinity is equal to \rev{the negative} lowest unoccupied KS eigenvalue, $-\varepsilon_{c}(1/2)$, with half-occupation. The band gap, $E_g$, therefore, is just the difference between these two \rev{half-occupied} eigenvalues. However, instead of performing calculations with half-occupied KS orbitals, in the LDA-1/2 method, a so-called ``self-energy potential'', $V_S$, is subtracted from the LDA XC potential, $v_{XC}$. The new KS potential provides eigenvalues $\varepsilon_v$ and $\varepsilon_c$ which are equal to those which would be obtained with the ``half-ionizing'' procedure. The local potential $V_S$ can be approximated from atomic calculations as a difference between the atomic KS potential of the neutral and the half-ionized atom, for each species in the crystal.\cite{Ferreira2008,Ferreira2011} This procedure implements the idea proposed by Slater of creating a localized hole which could act in solids in the same way as the ``half-ionizing'' technique works for atoms and molecules.\cite{Slater1971,Slater1972a,Slater1972b,Ferreira2008,Ferreira2011} The self-energy potential can indeed remedy the \rev{delocalization} error of charge densities in LDA, and therefore ameliorates the quality of KS eigenvalues.\cite{MoriSanchez2008} More details about the method itself and its derivation can be found in Refs. \onlinecite{Ferreira2008} and \onlinecite{Ferreira2011}.

\par After the (g)KS calculations, QP corrected eigenvalues $\varepsilon^{QP}_{n\mathbf{k}}$ are obtained in the framework of MBPT, within the \GW{} approximation, as follows:
\begin{equation}
\varepsilon^{QP}_{n\mathbf{k}} = \varepsilon_{n\mathbf{k}} + Z_{n\mathbf{k}}\langle \phi_{n\mathbf{k}}|\mathrm{Re}[\Sigma(\varepsilon_{n\mathbf{k}})] - v_{XC}|\phi_{n\mathbf{k}}\rangle.
\end{equation}
Here $\varepsilon_{n\mathbf{k}}$ means the (g)KS eigenvalue with Bloch vector $\mathbf{k}$ and band index $n$, $\phi_{n\mathbf{k}}$ is the corresponding (g)KS wavefunction. $Z_{n\mathbf{k}}$ stands for the QP renormalization factor. The self-energy operator $\Sigma$ is written as a convolution between the one-electron Green's function $G$ and the dynamically screened Coulomb interaction $W$. 

\par
\begin{figure}
\centering
\includegraphics[scale=1]{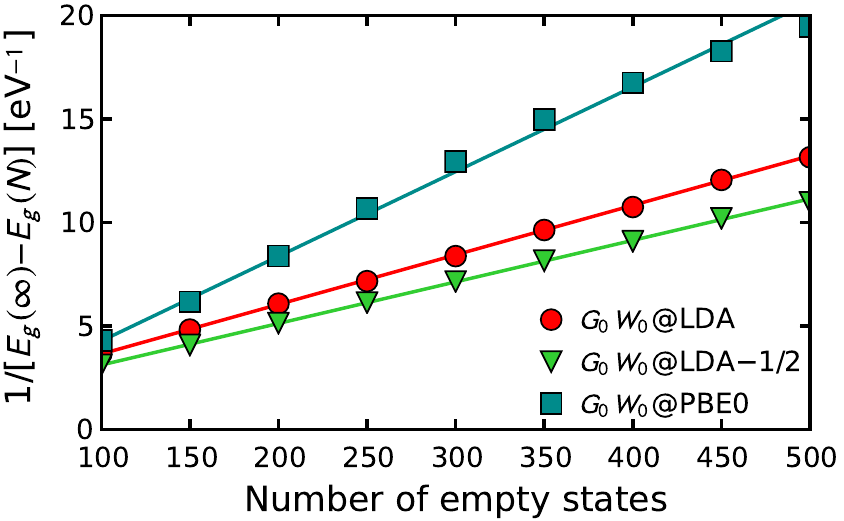}
\caption{Example of the extrapolation scheme for MgO. The extrapolated band gap $E_g(\infty)$ used in this figure \rev{is 7.638, 8.021, and 8.929 eV,} for \GW{} on top of LDA, LDA-1/2, and PBE0, respectively. All the calculations were \rev{carried out with a $4\times 4 \times 4$ $k$-grid}.
}\label{fig-extrapolation}
\end{figure}

 All the calculations are carried out using the all-electron full-potential computer package \exciting,\cite{Gulans2014} which implements the linearized augmented planewave (LAPW) method. 
\rev{Special care} is taken in \GW{} calculations with respect to the number of empty states and the number of $k$ points, following \rev{the procedure} proposed in Refs. \onlinecite{Nabok2016} and \onlinecite{Friedrich2011}. Local orbitals are included in our calculations in order to better  represent high-lying unoccupied states in the \GW{} approach.\cite{Nabok2016,Friedrich2011,Jiang2016} To keep the computational cost reasonable, we initially perform calculations with a \rev{$4 \times 4 \times 4$ $k$ grid to reach convergence} regarding the number of empty states $N$ (which is varied from 100 up to a maximum of 300 or 600, depending on the material).\cite{Nabok2016,Friedrich2011} The band gap is then \rev{extrapolated  according} to the expression\cite{Nabok2016,Friedrich2011}
\begin{equation}\label{eq.extrapolation}
E_g^{4\times 4 \times 4}(N) = E_{g}^{4\times 4 \times 4}(\infty)+\frac{A}{B+N},
\end{equation}
where $E_g^{4\times 4 \times 4}(\infty)$, \rev{$A$,} and $B$ are fit parameters. \rev{$E_g^{4\times 4 \times 4}(\infty)$ is} the band gap for an infinite number of empty states, i.e. the fully converged value. In Fig. \ref{fig-extrapolation}, we show an example of this extrapolation scheme for MgO. After obtaining $E_g^{4\times 4 \times 4}(\infty)$ by a non-linear least square fit, we plot $1/[E_g^{4\times 4 \times 4}(N) - E_{g}^{4\times 4 \times 4}(\infty)]$ as a function of the number of empty states. If the extrapolation proposed in Eq. (\ref{eq.extrapolation}) is obeyed, a linear behavior is expected, and this is truly what is observed in Fig. \ref{fig-extrapolation}. The same also holds for the other materials studied here.

\par In a second step, we extrapolate the band gap with respect to the $k$ points, as in Refs. \onlinecite{Nabok2016} and \onlinecite{Klimes2014}:
\begin{equation}
E_g^{QP} = E_{g}^{4\times 4 \times 4}(\infty)+E_{g}^{6\times 6 \times 6}(100)-E_{g}^{4\times 4 \times 4}(100),
\end{equation}
where $E_{g}^{6\times 6 \times 6}(100)$ is the \GW{} band gap calculated with 100 empty states and with a $k$ grid of $6\times6\times6$. We regard $E_g^{QP}$ as the \GW{} band gap, now extrapolated to an infinite basis. 

\par To benchmark the quality of the starting point, we choose the following solids: C, Si, SiC, AlP, LiF, MgO, Ne, Ar, GaN, GaAs, CdS, ZnS, and ZnO. In each case, and for each starting point, we obtain the band gap, the position of $d$ levels (in case they are present), and the band structure, comparing our calculations with experimental data. These materials are studied in zinc blende (zb) structure, except MgO and LiF (rock salt), and the noble-gas solids Ne and Ar (face centered cubic). For all systems, we employ experimental lattice parameters. 

\begin{table*}[htb]
\centering
\caption{\rev{Band gaps, in units of eV, for different XC functionals and with \GW{} on top, respectively. The} valence band maximum is at $\Gamma$. \rev{The conduction band minimum lies at the point indicated in parenthesis or between $\Gamma$ and $X$ otherwise.}
``Exp.'' refers to measured band gaps, while in ``Exp.*'' contributions coming from spin-orbit coupling and zero-point vibrations \rev{are subtracted}. For ZnO and GaN, the \rev{ZPE} corrections were taken from experimental data for the wurtzite phase.
Experimental data: 
\Ref{a}{Grueneis2014};
\Ref{b}{Haas2009};
\Ref{c}{Shishkin2007};
\Ref{d}{Schilfgaarde2006a};
\Ref{e}{Kotani2007};
\Ref{f}{Rohlfing1993};
\Ref{g}{Vurgaftman2001};
\Ref{h}{Bechstedt2009};
\Ref{i}{Vurgaftman2003};
\Ref{j}{Adachi};
\Ref{k}{Monserrat2014};
\Ref{l}{Giustino2010};
\Ref{m}{Cardona2005};
\Ref{n}{Antonius2015}.
}\label{tab-gaps}
\begin{tabular}{l|S|SSS|SSS|SSSS}
\hline
	&		&		&	{(g)KS}	&		&		&	{\GW{@}}	&		&		&		&		&		\\
	& {\emph{a} (\AA)}	&	{LDA}	&	{LDA-1/2}	&	{PBE0}	&	{LDA}	&	{LDA-1/2}	&	{PBE0}	&	{Exp.*}	&	{Exp.}	&	{\rev{ZPE}}	&	$\Delta_{SO}$	\\
\hline
C 	&	3.567$^a$	&	4.10	&	4.82	&	6.09	&	5.79	&	6.00	&	6.28	&	5.89	&	5.48$^d$	&	0.41$^k$	&	0.00$^d$	\\
C ($\Gamma$)&		&	5.55	&	5.87	&	7.75	&	7.43	&	7.61	&	8.10	&	7.74	&	7.14$^e$	&	0.6$^l$	&	0.00$^d$	\\
SiC (X)	&	4.340$^b$	&	1.31	&	2.26	&	2.97	&	2.43	&	2.57	&	3.13	&	2.50	&	2.39$^f$	&	0.11$^k$	&	0.00$^a$	\\
SiC ($\Gamma$)&		&	6.43	&	7.05	&	8.46	&	7.51	&	7.78	&	8.47	&	7.86	&	7.75$^f$	&	0.11$^k$	&	0.00$^a$	\\
Si 	&	5.431$^a$	&	0.48	&	1.09	&	1.70	&	1.14	&	1.36	&	1.40	&	1.23	&	1.17$^e$	&	0.05$^k$	&	0.01$^d$	\\
Si	($\Gamma$) &		&	2.53	&	2.79	&	4.00	&	3.23	&	3.35	&	3.46	&	3.41	&	3.35$^e$	&	0.05$^k$	&	0.01$^d$	\\
AlP (X)	&	5.463$^a$	&	1.45	&	2.82	&	2.97	&	2.38	&	2.48	&	3.03	&	2.56	&	2.52$^g$	&	0.02$^m$	&	0.02$^g$	\\
AlP	($\Gamma$) &		&	3.09	&	4.16	&	4.92	&	4.09	&	4.37	&	4.96	&		&		&		&		\\
LiF	($\Gamma$) &	4.010$^c$	&	8.94	&	12.44	&	12.27	&	14.10	&	14.64	&	15.47	&	14.48	&	14.20$^c$	&	0.28$^n$	&		\\
Ne	($\Gamma$) &	4.430$^c$	&	11.44	&	16.94	&	15.18	&	20.70	&	21.11	&	21.40	&	21.70	&	21.70$^c$	&		&		\\
Ar	($\Gamma$) &	5.260$^c$	&	8.18	&	11.52	&	11.13	&	13.28	&	14.05	&	14.35	&	14.20	&	14.20$^c$	&		&		\\
MgO	($\Gamma$) &	4.211$^a$	&	4.67	&	7.19	&	7.24	&	7.62	&	8.02	&	8.80	&	7.95	&	7.80$^a$	&	0.15$^m$	&	0.00$^a$	\\
ZnO	($\Gamma$) &	4.580$^c$	&	0.62	&	3.18	&	3.09	&	2.66	&	3.19	&	4.07	&	3.36	&	3.20$^h$	&	0.16$^m$	&	0.00$^a$	\\
GaN	($\Gamma$) &	4.52$^b$	&	1.68	&	3.53	&	3.64	&	3.04	&	3.22	&	3.71	&	3.47	&	3.30$^i$	&	0.17$^m$	&	0.00$^a$	\\
GaAs ($\Gamma$)	&	5.654$^a$	&	0.28	&	1.34	&	2.09	&	1.10	&	1.43	&	1.82	&	1.68	&	1.52$^g$	&	0.05$^m$	&	0.11$^d$	\\
ZnS	($\Gamma$)&	5.409$^a$	&	1.84	&	3.59	&	4.00	&	3.35	&	3.74	&	4.33	&	3.92	&	3.81$^j$	&	0.08$^m$	&	0.03$^d$	\\
CdS	($\Gamma$)&	5.818$^a$	&	0.88	&	2.69	&	2.84	&	2.04	&	2.47	&	2.88	&	2.53	&	2.43$^a$	&	0.07$^m$	&	0.03$^a$	\\ \hline
\end{tabular}
\end{table*}
\section{Results and Discussion}\label{sec-results}
\par Our calculated band gaps are presented in Table \ref{tab-gaps} \rev{and compared with experimental data from the literature. For this purpose,} we remove from the measured band gaps the contributions due to spin-orbit (SO) coupling and zero-point renormalization energy (\rev{ZPE}). \rev{Values obtained this way are marked by a star. Special attention} must be paid to the corrections arising from the \rev{ZPE}. It quantifies the electron-phonon coupling strength, and is specially important for semiconductors composed by light elements, such as those from the second period. For instance, for diamond the \rev{ZPE} is as high as 0.6 eV.\cite{Schilfgaarde2006a} The \rev{ZPE} has been pointed out in the literature as the most important source of discrepancies between experiments and calculations, which usually do not take this effect into account.\cite{Chen2014,Schilfgaarde2006b,Trani2010,Draxl2014,Klimes2014,Jain2014,Gerosa2015}
For the solids addressed here, only for GaAs, which contains the heaviest anion, spin-orbit coupling has a higher contribution to the renormalization of the band gap than the \rev{ZPE}.

\subsection{Band gaps from (generalized) Kohn-Sham calculations}\label{subsec-KS}

\par
\begin{figure*}[hbt]
\includegraphics[scale=1]{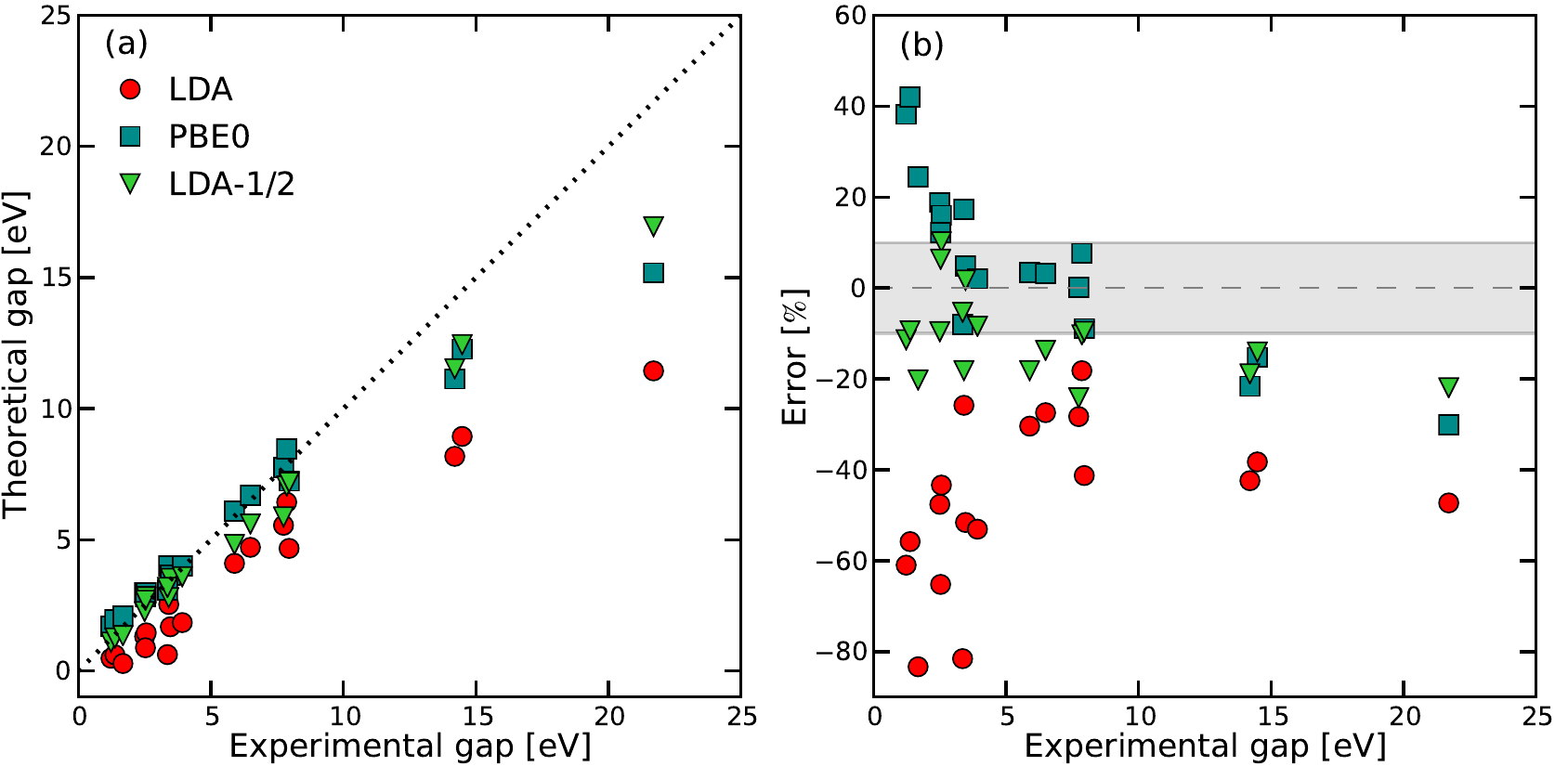}
\caption{(a): Theoretical band gaps obtained within (g)KS approaches compared with experiment. The dotted line indicates the desired result, where theory meets experiment. (b): Relative error in the band gap obtained by (g)KS calculations. The shaded area \rev{indicates} deviations within 10\%.}\label{fig-errorKS}
\end{figure*}

 In Fig. \ref{fig-errorKS} (a), we plot the (g)KS band gaps given in \rev{Table} \ref{tab-gaps}. The errors corresponding to each (g)KS approach \rev{are} presented in Fig. \ref{fig-errorKS} (b). Obviously, this figure reflects the typical underestimation by LDA, where semiconductors with $d$ electrons, such as ZnO, GaN, GaAs, ZnS, and CdS, appear among those with the highest discrepancy. The corresponding $d$ levels are not sufficiently localized in LDA and tend to be placed too high in energy, pushing the top of valence band (VB) upward.\cite{Furthmueller2005,Chen2014}

\par Employing PBE0 improves substantially, as calculated band gaps become considerably closer to experiment.\cite{Chen2014,Chen2012,Betzinger2010,Marques2011,Skone2014} However, the accuracy of PBE0 is highly related to the band gap itself -- as it can be verified in Fig. \ref{fig-errorKS} (b). For larger band gaps, an underestimation is observed, and the reverse occurs for narrower band gaps. This agrees with previous observations.\cite{Chen2012,Chen2014} According to our calculations, the optimal interval, with smallest errors, lies between $\sim$ 3 and 7 eV.

\par \rev{The LDA-1/2 method performs} considerably better than LDA, and its accuracy is comparable with PBE0. Although overall LDA-1/2 still tends to underestimate band gaps, similar to PBE0 for compounds with large band gaps, it fixes the underestimation present in LDA, enhancing substantially the agreement experiment.

\par A statistical measure of the quality of LDA, \rev{PBE0,} and LDA-1/2 in predicting band gaps is obtained by a linear fit, \rev{$y=\gamma x$}, of the data displayed in Fig. \ref{fig-errorKS} (a). The closer \rev{$\gamma$} is to 1, the \rev{better} the calculations can reproduce experimental data. In Table \ref{tab-fitgap}, we show the results of this linear fit. Naturally, LDA provides the worst agreement with experiment, while LDA-1/2 and PBE0 exhibit almost the same slope. From these preliminary observations, we can already anticipate that LDA-1/2 should be better than LDA as starting point for \GW{} calculations, and probably as satisfactory as PBE0.

\begin{table}
\caption{Parameters of a linear fit, \rev{$y=\gamma x$}, through the data plotted in Fig. \ref{fig-errorKS} (a) and in Fig. \ref{fig-errorGW} (a).}\label{tab-fitgap}
\begin{tabular}{L{30mm}C{25mm}}
\hline
 Method & $\gamma$ \\\hline
LDA & $0.59\pm0.02$\\
PBE0 & $0.83\pm0.03$\\
LDA-1/2 & $0.82\pm0.02$ \\
\GW{@}LDA & $0.951\pm0.007$ \\
\GW{@}PBE0 & $1.031\pm0.012$\\
\GW{@}LDA-1/2 & $0.987\pm0.005$ \\\hline
\end{tabular}
\end{table}

\subsection{Band gaps from \GW{} based on different starting points}\label{subsec-GW}
\par
\begin{figure*}[htb]
\includegraphics[scale=1]{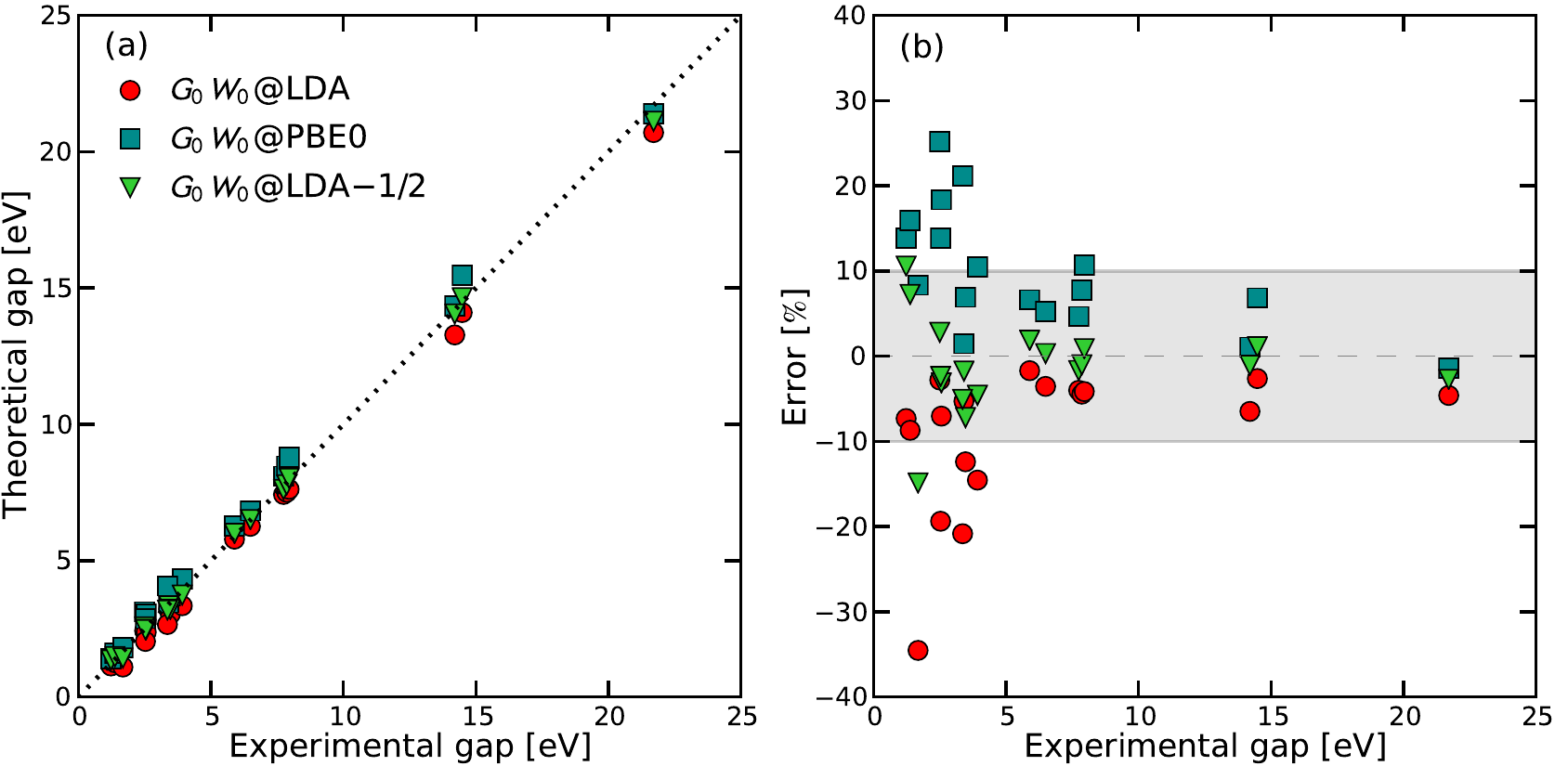}
\caption{Same as Fig. \ref{fig-errorKS}, but for the \GW{} approximation.}\label{fig-errorGW}
\end{figure*}

 In Fig. \ref{fig-errorGW} (a), we compare band gaps calculated within \rev{the} \GW{} approximation based on different reference (g)KS hamiltonians with experimental ones. The agreement between theory and experiment is better \rev{assessed by a} linear fit \rev{$y=\gamma x$ through} the points presented in Fig. \ref{fig-errorGW}. Table \ref{tab-fitgap} displays the \rev{corresponding slopes}. \rev{The best} agreement with experiment is reached when LDA-1/2 is employed as starting point for \GW{}. We shall resume this.

\par \rev{Figure} \ref{fig-errorGW} (b) shows the errors with respect to experiments, reflecting the \rev{well-known starting-point dependence\cite{Koerzdoerfer2012,Chen2014,Gallandi2016,Marom2012,Pinheiro2015,Blase2011,Faber2011,
Marom2011,Ren2012,Bruneval2013} in} a quantitative way.
Although the improvement over LDA is already very impressive, \GW{@}LDA still systematically underestimates band gaps, as already observed in the literature.\cite{Bechstedt2009,Schilfgaarde2006b,Kotani2007,Fuchs2007} On the other hand, \GW{@}PBE0 overestimates them, Ne being an exception. This agrees with \rev{previous investigations,}\cite{Chen2014,Isseroff2012} although for molecules, results of \GW{@}PBE0 for the highest occupied molecular orbital show much better agreement with experiments.\cite{Bruneval2013,Valentin} Like for PBE0, the accuracy of \GW{@}PBE0 highly correlates with the band gap: the larger the band gap, the better \rev{it is}. In comparison with \GW{@}LDA, there is no overall advantage of \GW{@}PBE0. However, the latter is in better agreement with experiments for materials with $d$ electrons, i.e., ZnO, GaN, GaAs, ZnS, CdS. Nevertheless, for these materials, except for GaAs, band gaps obtained with PBE0 are even better than the ones obtained with \GW{@}PBE0. 

\par \rev{Figure} \ref{fig-errorGW} (b) also shows that the accuracy of \GW{@}LDA-1/2 surpasses both \GW{@}LDA and \GW{@}PBE0 for almost all the materials. ZnO is a particular case that attracts our attention. This oxide has been a long-standing issue of \GW{} calculations on top of semi-local functionals. \cite{Chen2014,Friedrich2011,Stankovski2011,Klimes2014,Lim2012} The band gap obtained \rev{with} LDA-1/2, 3.18 eV, is in much better agreement with \rev{the} experimental gap \rev{of} 3.36 eV, than the \rev{one} calculated \rev{with} LDA, 0.62 eV. \rev{While \GW{@}LDA leads to a band gap of 2.66 eV, which is still substantially underestimated, \GW{@}LDA-1/2 practically does not alter the value obtained with its starting point. The situation is different for PBE0 and \GW{@}PBE0. The band gap obtained with PBE0, 3.09 eV, is already very close to the experimental one, but \GW{@}PBE0 deteriorates this result by overestimating it.} 

\par
\begin{figure}[htb]
\includegraphics[scale=1]{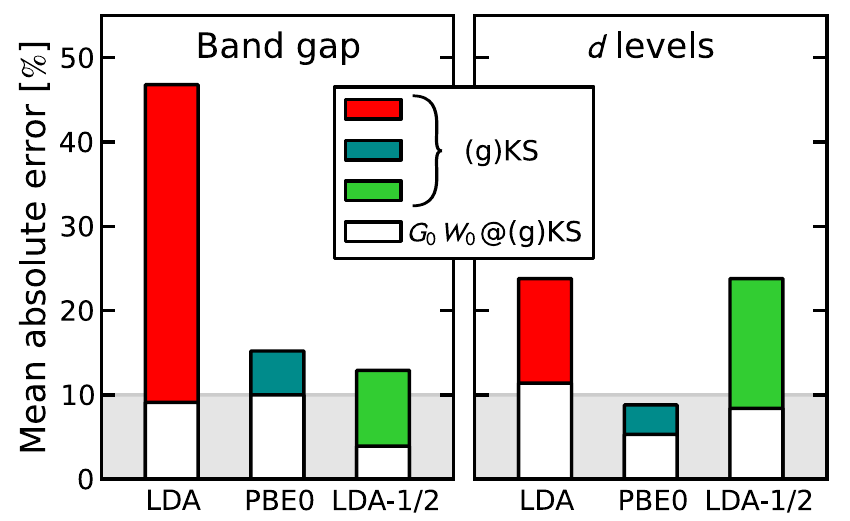}
\caption{Overview of how the MAE is decreased after introducing QP corrections. On the left side, band gaps, and on the right side, $d$ levels. An error bar of 10\% is depicted in gray in the background.}
\label{fig-bar}
\end{figure}

 \rev{Figure} \ref{fig-bar} depicts the mean \rev{absolute} error (MAE), giving an overview of the accuracy reached by each method. PBE0 and LDA-1/2 have a MAE of 15.2\% and 12.9\%, respectively, which \rev{are} less than \rev{half} of the MAE of LDA, 46.8\%. The LDA-1/2 method itself, despite being a method with a local KS potential, has a MAE comparable to \GW{@}LDA(9.1\%) and \GW{@}PBE0 (10.0\%). It is impressive that the MAE of \GW{@}LDA-1/2, of 3.9\%, is less than half of the values of \GW{@}LDA and \GW{@}PBE0.

\par
\begin{figure}[htb]
\includegraphics[scale=1]{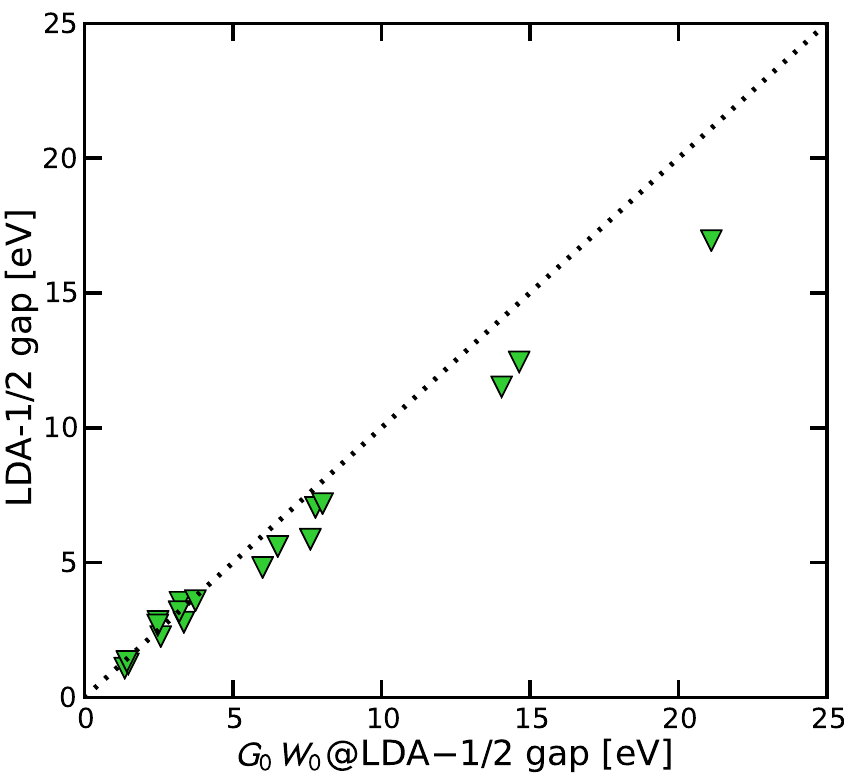}
\caption{Comparison between band gaps obtained with LDA-1/2 and \GW{@}LDA-1/2. The dotted line shows the ideal case, where LDA-1/2 could replace \GW{@}LDA-1/2.}\label{fig-LDA05}
\end{figure}

 We further address the accuracy of LDA-1/2 in Fig. \ref{fig-LDA05}, by comparing band gaps obtained with LDA-1/2 and \GW{@}LDA-1/2. It is interesting to know the cases when \rev{one} can avoid the \GW{} step. \rev{Figure} \ref{fig-LDA05} suggests that LDA-1/2 agrees \rev{better} with \GW{@}LDA-1/2 for small band gaps. To quantify this assessment, a linear fit \rev{$y=\gamma x$} is performed. For band gaps ranging from 0 to 10 eV, \rev{$\gamma$} is found to be $0.88\pm0.02$, closer to 1, when compared to the range 0 to 25 eV, when $a$ is equal to $0.84\pm0.02$. This means that for compounds with band gaps \rev{between 0 and 10 eV}, \rev{calculations with LDA-1/2 lead to band gaps which are} on average roughly 90\% of \rev{the values} obtained with \GW{@}LDA-1/2. This provides a measure of the extent up to which LDA-1/2 can approximate \GW{@}LDA-1/2.

\subsection{\textit{d}-band positions}
\par
\begin{figure}[htb]
\includegraphics[scale=1]{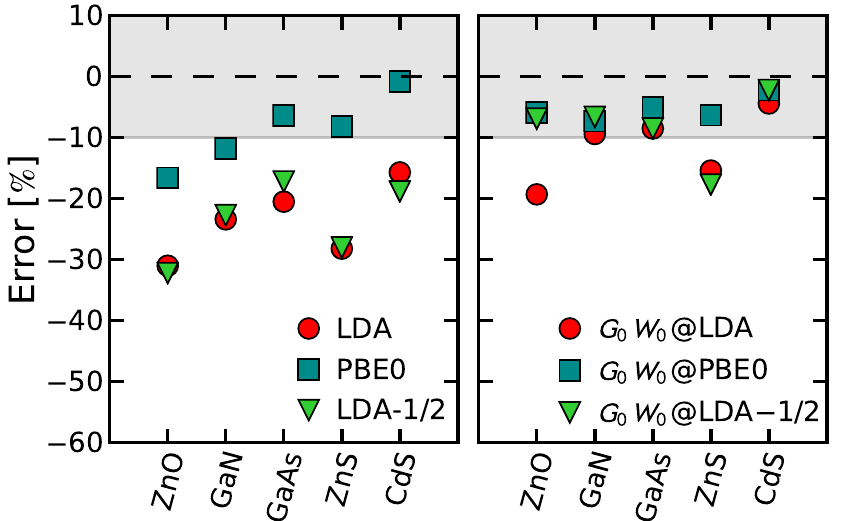}
\caption{Relative error of calculated $d$-band positions, compared to experiments. On the left side, results obtained within a (g)KS framework; on the right side, \rev{calculations including} QP corrections. Experimental data were taken from Refs.  \onlinecite{Goepel1982} and \onlinecite{Vogel1995} (wurtzite ZnO), Ref. \onlinecite{Ding1997} (GaN), Ref. \onlinecite{Ley1974} (GaAs and ZnS), Ref. \onlinecite{Stampfl1997} (CdS).
}\label{fig-error-dlevel}
\end{figure}

 Among the solids chosen for our benchmark, only ZnO, GaN, GaAs, ZnS, and CdS have $d$ bands. \rev{Whereas} ZnO, \rev{ZnS,} and CdS exhibit shallow $d$ states located 7-9 eV below the VB maximum (VBM),\cite{Goepel1982,Vogel1995,Ley1974,Stampfl1997} \rev{they are deeper in GaN and GaAs,} i.e., 17-19 eV below the VBM.\cite{Ley1974,Ding1997} We determine the position of these bands as an average among the corresponding eigenvalues at the $\Gamma$ point, and \rev{compare} our results with measurements reported in the literature. In Fig. \ref{fig-error-dlevel}, we display the \rev{differences between} calculations and experiments. A negative (positive) value means that the respective method places the $d$ states too high (low) in energy. 

\par No substantial differences are observed between LDA and LDA-1/2. PBE0, in contrast, places $d$ levels at lower energies than LDA and LDA-1/2. The agreement with experiments is clearly better, due to the inclusion of \rev{Hartree-Fock exchange}, as expected also from a previous investigation with the exact exchange functional.\cite{Sharma2005} 

\par QP corrections on the \GW{} level push the position of $d$ bands down when compared to the underlying starting point, improving the agreement with experiments. This is \rev{evident from} Fig. \ref{fig-bar}, where the MAE is depicted. \rev{\GW{@}PBE0 predicts the positions of $d$ states with the highest accuracy, in other words, PBE0 represents for this purpose the best starting point for \GW{}.} The improvement of \GW{@}LDA and \GW{}@LDA-1/2 over their respective starting points is notable. Although $d$ levels are wrongly positioned by both LDA and LDA-1/2, only \GW{@}LDA-1/2 can cure this deficiency. We attribute this to the \rev{more localized} LDA-1/2 wavefunctions, \rev{caused by the attractive potential which LDA-1/2 introduces in the KS equations.} This helps to heal the delocalization error present in LDA.

\par The fact that $d$ states obtained with LDA and LDA-1/2 are not different is not surprising.  The LDA-1/2 method \rev{employs Slater's} transition state technique to improve the description of the VBM and \rev{conduction band minimum (CBM), and the results from sections} \ref{subsec-KS} and \ref{subsec-GW} confirm that this goal is really achieved. \rev{Conversely}, for states far apart from the VBM and CBM, such as the $d$ levels we are considering here, LDA-1/2 is not expected to have the same efficiency simply because it was not designed to do so. \rev{However}, we stress that, although the position of $d$ levels obtained with LDA and LDA-1/2 is basically the same, \GW{@}LDA-1/2 is \rev{more} accurate than \GW{@}LDA. 
\rev{While the MAE of \GW{@}LDA is} 11.4\%, the MAE of \GW{@}LDA-1/2 is 8.4\%.

\par At this point, it is worthwhile to comment about ZnO. The position of \rev{the} $d$ levels is wrongly predicted by LDA with a relative error of more than 30\%, the largest one among all the 5 compounds with $d$ electrons addressed in this paper. \GW{@}LDA, although improving over LDA, still places $d$ bands incorrectly with a relatively high error of \rev{20\%}. PBE0 gives rise to better results than LDA, even though the relative error still remains as high as 17\%. In \GW{@}PBE0, the error is decreased to \rev{6\%}. Finally, while the LDA-1/2 method \rev{exhibits} the same error as LDA, in \GW{@}LDA-1/2, the position of $d$ levels is in better agreement with the \rev{experimental one.}

\subsection{Band structures}
\begin{figure}[htb]
\centering
\includegraphics[scale=1]{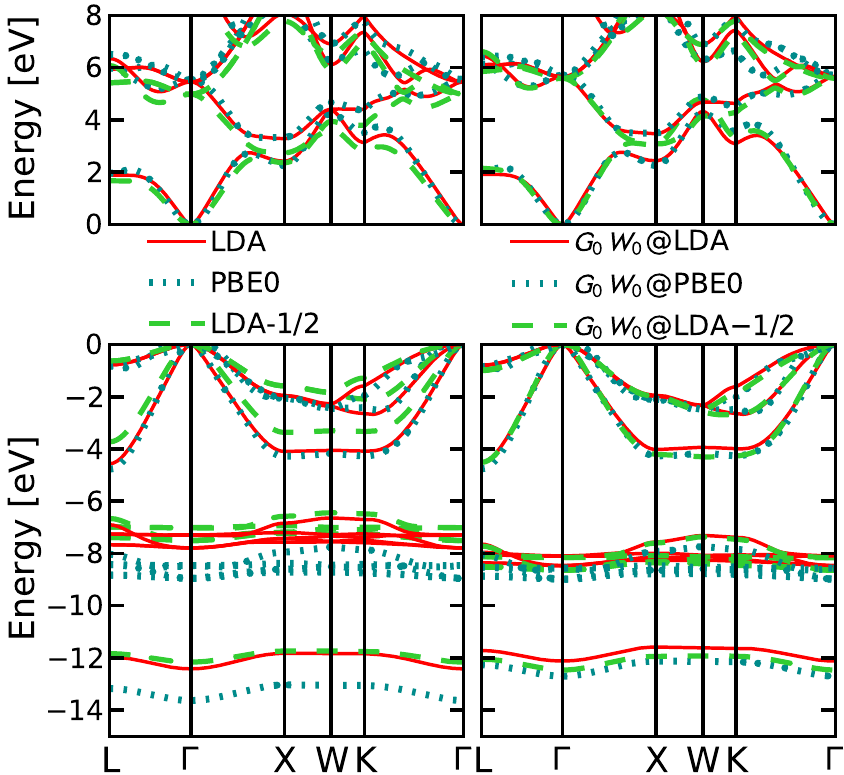}
\caption{Band structure of CdS. Calculations within (g)KS schemes on the left, and including \GW{} corrections on the right. At the top (bottom), unoccupied (occupied) bands are shown, with the \rev{CBM} (VBM) placed at zero.}
\label{fig-CdS-band2}
\end{figure}

\par
\begin{figure}[htb]
\centering
\includegraphics[scale=1]{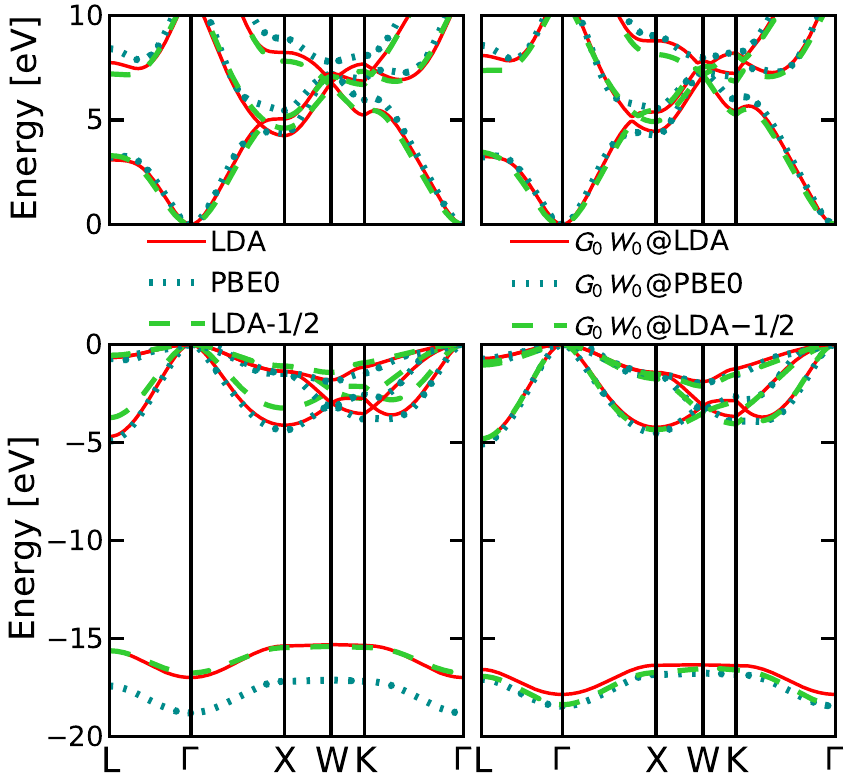}
\caption{Same as Fig. \ref{fig-CdS-band2}, but for MgO.}
\label{fig-MgO}
\end{figure}

 As a final point, we address the \rev{starting point effect on} the band structures obtained within \GW{}. CdS and MgO are chosen as examples. \rev{We} highlight the differences among the band structures obtained within LDA, PBE0, and LDA-1/2, as well as among \GW{} \rev{results for the case of CdS in Fig. \ref{fig-CdS-band2} and for MgO in Fig. \ref{fig-MgO}}.

\par \rev{From Figs. \ref{fig-CdS-band2} and \ref{fig-MgO}}, we see that, \rev{considering CdS (or MgO, accordingly),} the shape of the VB calculated within LDA and PBE0 is essentially the same, and is not altered by QP corrections on the \GW{} level. The LDA-1/2 method, in turn, deforms the VB, decreasing its bandwidth \rev{in 0.8 and 1.2 eV for CdS and MgO, respectively}. The right shape is recovered by \GW{@}LDA-1/2, which agrees with \GW{@LDA} and \GW{@}PBE0. 

\par Even though \rev{the} $d$ levels of cadmium are better obtained with PBE0 than with LDA and LDA-1/2, the  semicore $3s$ state of sulfur is better described by the latter methods. \rev{As can be seen in Fig. \ref{fig-CdS-band2}, LDA and LDA-1/2 place these states around $-12$ eV, in agreement with the position obtained with \GW{}}. \rev{In the case of MgO, a similar situation occurs for the $2s$ states of oxygen. LDA and LDA-1/2 place these levels between $-15.6$ and $-17.0$ eV, while PBE0 places them between $-17.2$ and $-18.8$ eV. \GW{} bands agree with the position obtained by LDA and LDA-1/2.} 


\par The shape of the CB is similar in all three (g)KS schemes. LDA-1/2 does not diminish its band width as much as for the VB. 



\section{Conclusions}\label{sec-conclusions}
\par In this work, we propose and investigate a new starting point for \GW{} calculations -- the LDA-1/2 method. To benchmark the performance of LDA-1/2, also LDA and PBE0 \rev{have} been examined as starting points.
For \rev{the} band gaps \rev{of} the investigated materials, \GW{@}LDA-1/2 turns out to be the most accurate approach, leading to predictions in very nice agreement with experiments. In the case of $d$ bands, PBE0 proves to be the best starting point \rev{for} \GW{} calculations. Anyway, \GW{@}LDA-1/2 results are closer to experiment when compared to \GW{@}LDA. \rev{Especially} for materials with a small band gap, the LDA-1/2 method alone, i.e., without a subsequent \GW{} calculation, achieves considerable accuracy if one is interested in the band gap only. This scheme can be useful when studying more complex materials, such as interfaces, surfaces, \rev{or} heterostructures, as \GW{} may come with a prohibitive computational cost in these cases.

\section{Acknowledgments}
\par Support received from Alexander von-Humboldt Stiftung and Coordena\c{c}\~{a}o de Aperfei\c{c}oamento de Pessoal de N\'{i}vel Superior (CAPES) is thankfully acknowledged. We thank Nora Illanes Salas and Benjamin Aurich for critical reading of the manuscript.

\bibliography{References}

\end{document}